\documentclass[prc,aps,onecolumn,showpacs,preprint]{revtex4}
\usepackage{graphicx}
\usepackage{amsmath}

\begin{document}

\title{Single-energy amplitudes for pion photoproduction
in the first resonance region}

\author{
R.\ L.\ Workman}
\affiliation{
Center for Nuclear Studies,
Department of Physics\\
The George Washington University,
Washington, D.C. 20052}

\date{\today}

\begin{abstract}

We consider multipole amplitudes for low-energy pion photoproduction,
constructed with minimal model dependence, at single energies.
Comparisons with fits to the full resonance
region are made. Explanations are suggested for the discrepancies and 
further experiments are motivated.

\end{abstract}


\maketitle

\section{Introduction and Motivation}

In a series of papers, Grushin and collaborators~\cite{grushin}
extracted multipole amplitudes for $\pi^+ n$ photoproduction, 
$\pi^0 p$ photoproduction, and combined these to produce 
isospin components, from 280 MeV to 420 MeV, without employing
Watson's theorem. 
A number of subsequent studies~\cite{mukh} took this set to be the
least biased determination of multipoles over the delta resonance 
region. As
the amplitudes were obtained in the early 80's, prior to a number
of recent high-precision measurements, we have reexamined
these results and the methods used in their determination. 

Apart from checking old values, this exercise is relevant to
experimental programs now measuring complete, or nearly complete,
experiments for pion and kaon photoproduction. The relative
model-independence of this method allows checks of database consistency
which we will use to suggest further measurements. Below we also briefly 
compare the methods associated with amplitude reconstruction and
multipole fitting.

In Ref.\cite{grushin}, multipoles were extracted from
$\pi^+ n$ photoproduction data of type $S$ only. In the language of
Ref.~\cite{bds}, type-$S$ data include the unpolarized cross section
and single-polarization asymmetries  
($P$, $\Sigma$, $T$ ). As these do not constitute a complete
experiment, in the strict sense of Ref.~\cite{tabakin}, some assumptions
are required. The fits were performed between 280 MeV and 420 MeV, 
using a truncated multipole expansion,
including $E_{0+}$, $M_{1-}$, $E_{1+}$, and $M_{1+}$, the remaining 
terms assumed to be real and given by the electric Born terms. 

The multipoles and helicity amplitudes are related by
\begin{eqnarray}
H_1 &=& {\frac{1}{\sqrt{2}}} \cos {\theta \over 2} \sin \theta \sum_{\ell = 1}^{\infty} 
\left[ E_{\ell +} - M_{\ell +} - E_{(\ell + 1) - } 
-  M_{(\ell + 1) -} \right] \left( P_{\ell}^{''} - P_{\ell + 1}^{''} \right), \nonumber \\
H_2 &=& {\frac{1}{\sqrt{2}}} \cos {\theta \over 2} \sum_{\ell = 0}^{\infty} 
\left[ (\ell + 2)E_{\ell +} + \ell M_{\ell +} + \ell E_{(\ell + 1) - } 
- (\ell +2) M_{(\ell + 1) -} \right] \left( P_{\ell}^{'} - P_{\ell + 1}^{'} \right), \nonumber \\
H_3 &=& {\frac{1}{\sqrt{2}}} \sin {\theta \over 2} \sin \theta \sum_{\ell = 1}^{\infty} 
\left[ (E_{\ell +} - M_{\ell +} +  E_{(\ell + 1) - } 
+  M_{(\ell + 1) -} \right] \left( P_{\ell}^{''} + P_{\ell + 1}^{''} \right), \nonumber \\
H_4 &=& {\frac{1}{\sqrt{2}}} \sin {\theta \over 2} \sum_{\ell = 0}^{\infty} 
\left[ (\ell + 2)E_{\ell +} + \ell M_{\ell +} - \ell E_{(\ell + 1) - } 
+ (\ell +2) M_{(\ell + 1) -} \right] \left( P_{\ell}^{'} + P_{\ell + 1}^{'} \right) .
\end{eqnarray}
From these one can construct the transversity amplitudes~\cite{bds}, 
\begin{eqnarray}
b_1 &=& \frac{1}{2} \left[ \left( H_1 + H_4 \right) \; + \; i \; \left( H_2 - H_3 \right) \right] ,
\nonumber \\
b_2 &=& \frac{1}{2} \left[ \left( H_1 + H_4 \right) \; - \; i \; \left( H_2 - H_3 \right) \right] ,
\nonumber \\
b_3 &=& \frac{1}{2} \left[ \left( H_1 - H_4 \right) \; - \; i \; \left( H_2 + H_3 \right) \right] ,
\nonumber \\
b_4 &=& \frac{1}{2} \left[ \left( H_1 - H_4 \right) \; + \; i \; \left( H_2 + H_3 \right) \right] ,
\end{eqnarray} 
which simplify the discussion of amplitude reconstruction, as the type-$S$ observables determine
their moduli
\begin{eqnarray}
\frac{d\sigma}{dt} &=& |b_1|^2 + |b_2|^2 + |b_3|^2 + |b_4|^2 , \nonumber \\
P \frac{d\sigma}{dt} &=& |b_1|^2 - |b_2|^2 + |b_3|^2 - |b_4|^2 , \nonumber \\
\Sigma \frac{d\sigma}{dt} &=& |b_1|^2 + |b_2|^2 - |b_3|^2 - |b_4|^2 , \nonumber \\
T \frac{d\sigma}{dt} &=& |b_1|^2 - |b_2|^2 - |b_3|^2 + |b_4|^2 .  
\end{eqnarray}
For $\pi^+ n$ photoproduction, 
the interference between (complex) fitted multipoles and a given (real)
high-$\ell$ contribution fixes the overall phase between transversity
amplitudes~\cite{bds}.  An amplitude reconstruction requires more
observables~\cite{tabakin}, and is the most model-independent method, but results in
transversity amplitudes, for each energy-angle pair, only up to an
unknown phase. If multipoles are the goal, an angular integral is required,
and this cannot be performed without determining the phase. Therefore, at
some point, every multipole analysis requires constraints beyond the
experimental data. 

\section{Fitting $\pi^+ n$ Data}

In order to check the results of Ref.~\cite{grushin}, 
data from
280 MeV to 420 MeV were fitted using the above prescription and
a more recent database~\cite{data,pipnew,beck}. 
The higher-$\ell$ multipoles were taken from the MAID analysis~\cite{MAID}, which includes 
vector-meson exchange, rather than a simple electric Born term. This modification
had a negligible effect on the fits.  The fitted multipoles were then compared to
the original determinations of Ref.~\cite{grushin} and single-energy
solutions (SES) tied to the SAID energy-dependent multipole analysis~\cite{SAID}.

\begin{table}
\begin{tabular*}{0.75\textwidth}{@{\extracolsep{\fill}}ccccc}
Multipole & Grushin~\cite{grushin} & SES & Fit1 & Fit2  \\
\hline
Re $E_{0+}$ & 17.18(0.29) & 16.2 & 16.72(0.18) & 16.17(0.23) \\
Im $E_{0+}$ & -3.10(0.98) & 0.57 & -3.41(0.87)  & 0.5  \\
Re $M_{1-}$ &  3.84(0.19) & 3.46 & 3.74(0.18) & 3.75(0.29) \\
Im $M_{1-}$ & -0.70(0.84) & -0.13 & -2.02(0.87) & 0.33(0.58) \\
Re $E_{1+}$ &  2.64(0.08) & 2.96 & 2.99(0.06) & 2.70(0.11)  \\
Im $E_{1+}$ &  0.00(0.26) & 0.70 & -0.08(0.29) & 0.78(0.19) \\
Re $M_{1+}$ & -16.00(0.30) & -14.85 & -16.24(0.24) & -14.76(0.18) \\
Im $M_{1+}$ & -6.76(1.10) & -9.63 & -5.96(0.98) & -10.06(0.35) \\
\end{tabular*}
\caption{\label{tab:pole3}Single-energy fits to $\pi^+ n$ data at
280 MeV (see text). Multipoles given in $10^{-3}/ m_{\pi}$ units.}
\end{table}
\begin{figure}
\includegraphics[ width=350pt, keepaspectratio, angle=90]{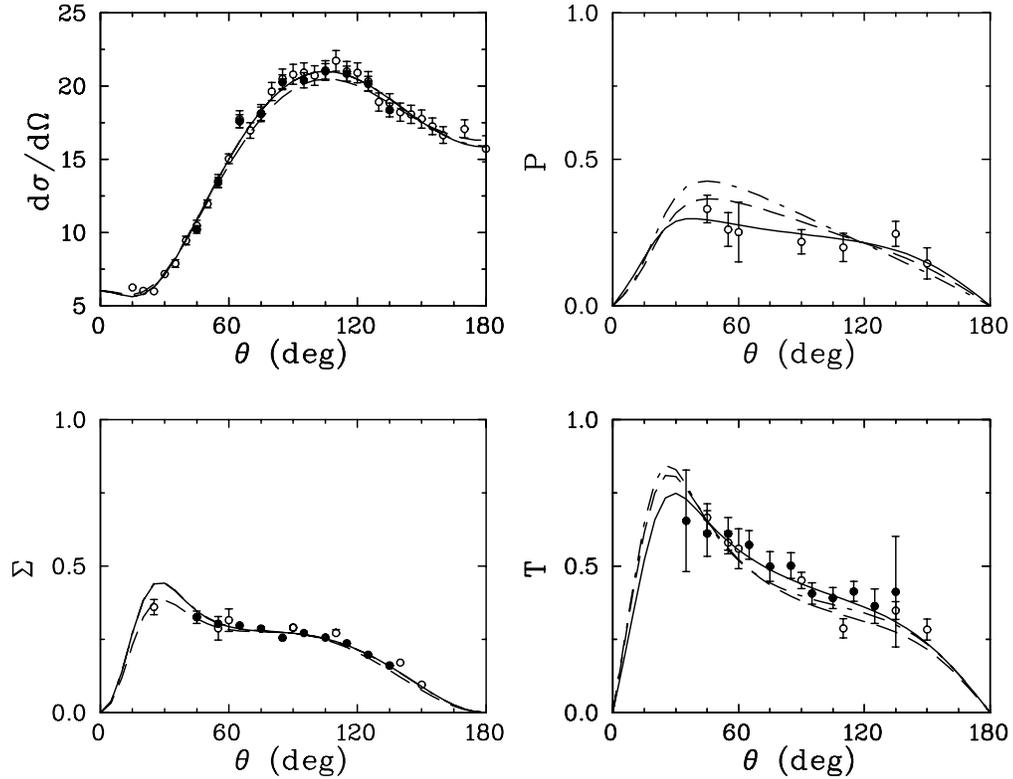}
\caption{\label{fig:g28}Fits to $\pi^+ n$ type-$S$ observables at
280 MeV. Fit1 (solid), SES (dashed), Fit2 (dot-dashed).
Post-1990 data~\cite{beck,pipnew} (solid), pre-1990 data~\cite{data} (open) symbols.}
\end{figure}

The present and original fits of Ref.~\cite{grushin} were generally consistent, except
in cases where more recent data contradicted older measurements. However, some very
large deviations from the SAID SES values were found at the lowest energy and at the
resonance energy (340 MeV). Comparisons are given in Tables I and II. The SES values
were obtained assuming Watson's theorem and fitting both neutral and charged pion 
data over narrow energy bins, assuming a linear energy dependence given by the 
energy-dependent fit. Errors on the fitted isospin
multipoles were generally in the 2$-$5\% range. 

The 280 MeV fit (Fit1) deviates from the trend shown in the 300 MeV to 420 MeV
results, and this was noticed in Ref.~\cite{grushin} where an inconsistency in the
data was suggested. The large negative fitted value for Im $E_{0+}$ at this energy contradicts
results, ($0.4\pm 0.2$) $10^{-3}/ m_{\pi}$, found in the SAID~\cite{SAID}, 
MAID~\cite{MAID}, and Bonn-Gatchina~\cite{BoGa} fits. 
As a test, this parameter was fixed and the remaining multipoles varied. The result (Fit2)
is consistent with the SES and is plotted, along with Fit1 and the SES, in Fig.~1. Note
that the modified value for Im $E_{0+}$ has an effect 
noticeable mainly in the recoil polarization,
the remaining quantities having been remeasured with greater precision. 

\begin{table}
\begin{tabular*}{0.75\textwidth}{@{\extracolsep{\fill}}ccccc}
Multipole & Grushin~\cite{grushin} & SES & Fit1 & Fit2  \\
\hline
Re $E_{0+}$ & 10.29(0.42) & 11.36 & 11.19(0.41) & 12.42(0.30) \\
Im $E_{0+}$ & 2.00(0.52) & -0.14 & 2.15(0.55)  & 0.0  \\
Re $M_{1-}$ &  1.82(1.40) & 4.53 & 2.89(1.45) & 4.32(1.27) \\
Im $M_{1-}$ & -0.11(0.22) & -0.17 & 1.17(0.30) & 0.50(0.31) \\
Re $E_{1+}$ &  0.30(0.35) & 1.79 & 0.69(0.38) & 1.22(0.29)  \\
Im $E_{1+}$ &  -0.41(0.10) & 0.30 &  0.47(0.14) & 0.18(0.16) \\
Re $M_{1+}$ & 1.34(0.98) & -1.82 & 1.11(0.24) & -1.66(0.61) \\
Im $M_{1+}$ & -19.26(0.46) & -18.29 & -18.84(0.22) & -18.31(0.21) \\
\end{tabular*}
\caption{\label{tab:m340}Single-energy fits to $\pi^+ n$ data at
340 MeV (see text). Multipoles given in $10^{-3}/ m_{\pi}$ units.}
\end{table}

\begin{figure}
\includegraphics[ width=350pt, keepaspectratio, angle=90]{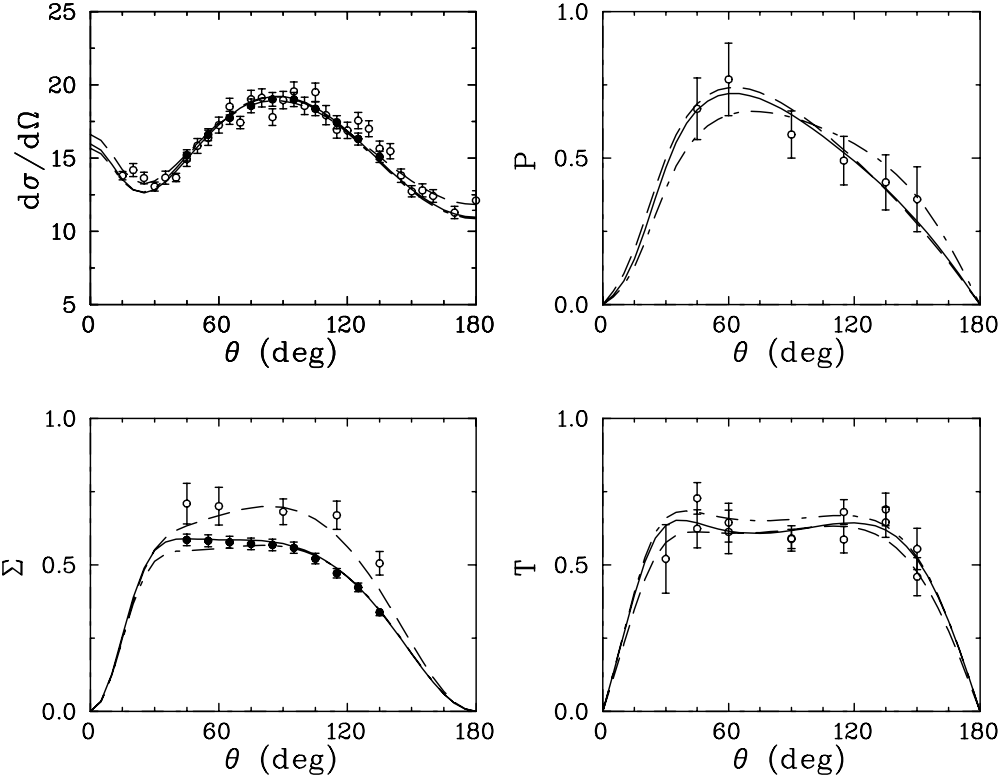}
\caption{\label{fig:g34} Fits to $\pi^+ n$ type-$S$ observables at
340 MeV. Fit1 (solid), Ref.~\cite{grushin} (dashed), Fit2 (dot-dashed).
Data as in Fig. 1.
}
\end{figure}

In Table II, a similar comparison is made at 340 MeV. In this case, however,
a precise remeasurement~\cite{beck} of $\Sigma$ found values shifted from the
set available to Grushin~\cite{grushin}. As a result, the refit (Fit 1) did not confirm
the original set of multipoles. Here too, a large value for Im $E_{0+}$ was found,
contradicting the SAID~\cite{SAID}, MAID~\cite{MAID},
and Bonn-Gatchina~\cite{BoGa} results, ($0\pm 0.4$) $10^{-3}/ m_{\pi}$.
Again, fixing this parameter to zero and refitting the remaining multipoles resulted in
a solution (Fit2) more compatible with the SES result. In Fig.~2 this readjustment
is expressed mainly in a different shape for $P$, which has sizeable error bars.

In summary, consistency between the SES results and the method of Ref.~\cite{grushin} is
sensitive to the rather poorly determined $P$ data.
More precise $P$ data would test the assumptions used in Ref.~\cite{grushin}. It should
be realized that almost every existing fit assumes the high$-\ell$ multipoles are real
and given by the Born plus vector meson exchange terms.
Predictions for the beam-target observables~\cite{bds}, given by Fit2, are compared to
SAID and available $G$ data~\cite{gdat} in Fig.~3.

\begin{figure}
\includegraphics[ width=350pt, keepaspectratio, angle=90]{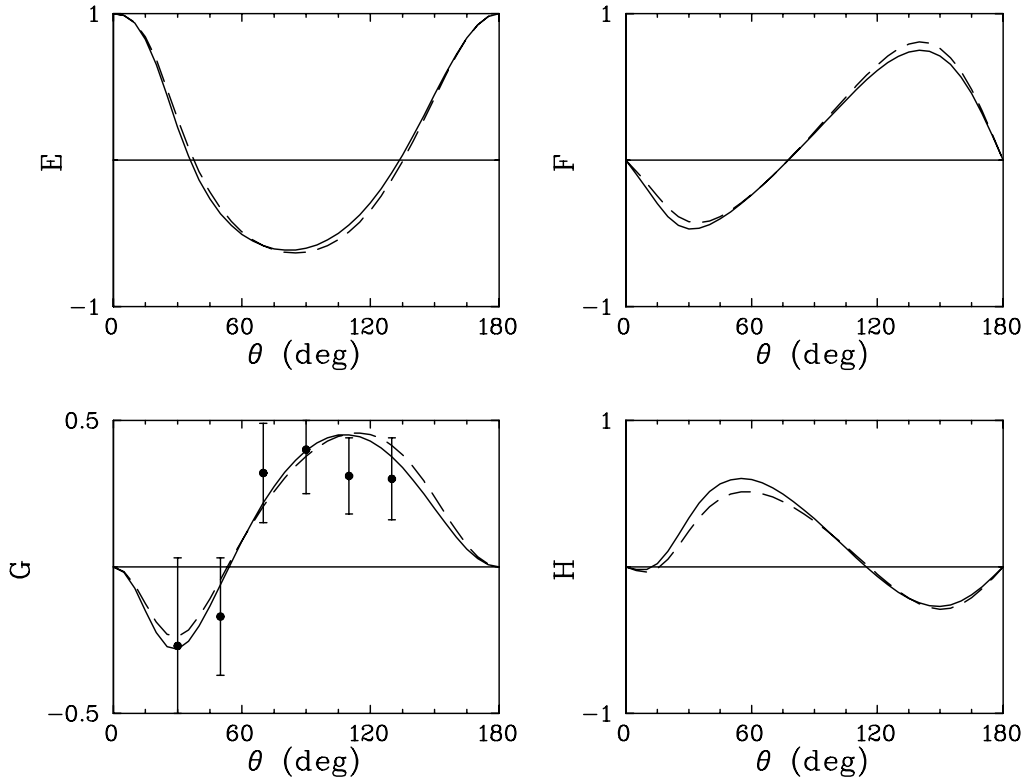}
\caption{\label{fig:gef} Prediction of $\pi^+n$ beam-target observables at
340 MeV from
Fit2 (solid) compared to the SAID energy-dependent fit.
Data from Ref.~\cite{gdat}.
}
\end{figure}

\section{Fitting $\pi^0 p$ Data}

If $\pi^+ n$ multipoles are available, they can be used to perform a similarly model
independent fit to $\pi^0 p$ photoproduction data. Unfortunately, the existing $P$
data for this channel are even worse over the delta resonance region. The set at
350 MeV has the clearest trend and has been fitted, again assuming a truncated multipole
expansion, ignoring higher$-\ell$ terms. Results are compared in Table III. 

\begin{table}
\begin{tabular*}{0.75\textwidth}{@{\extracolsep{\fill}}cccccc}
Multipole & Grushin~\cite{grushin} & SES & Fit1 & Fit2 & Fit3   \\
\hline
Re $E_{0+}$ & -1.64(0.46)  & -2.69  & -2.33(0.46) & -1.58(0.42) &  -1.20 \\
Im $E_{0+}$ & 1.03(0.24)   & 2.81   & 1.27(0.24)  &  2.14(0.31) &  2.36 \\
Re $M_{1-}$ &  -2.97(1.99) & -2.89  & -2.84(1.84) & -2.73(1.85)  &  18.67 \\
Im $M_{1-}$ &  0.57(0.17)  &  0.51  & -0.33(0.45) &  0.90(0.40)  &  -4.41 \\
Re $E_{1+}$ &  0.70(0.62)  & 1.34   & 0.63(0.58)  &  0.38(0.57) &  -7.74 \\
Im $E_{1+}$ &  -0.78(0.08) & -0.30  & -0.47(0.14) & -0.70(0.15)  &   1.44 \\
Re $M_{1+}$ &      -1.3    & -5.70  & -6.41(0.40) & -4.13        &  15.50 \\
Im $M_{1+}$ &  23.89(0.10) & 22.81  &   23.0      &  23.56      &  -6.36 \\
\end{tabular*}
\caption{\label{tab:m350}Single-energy fits to $\pi^0 p$ data at
350 MeV (see text). Multipoles given in $10^{-3}/ m_{\pi}$ units.}
\end{table}

\begin{figure}
\includegraphics[ width=350pt, keepaspectratio, angle=90]{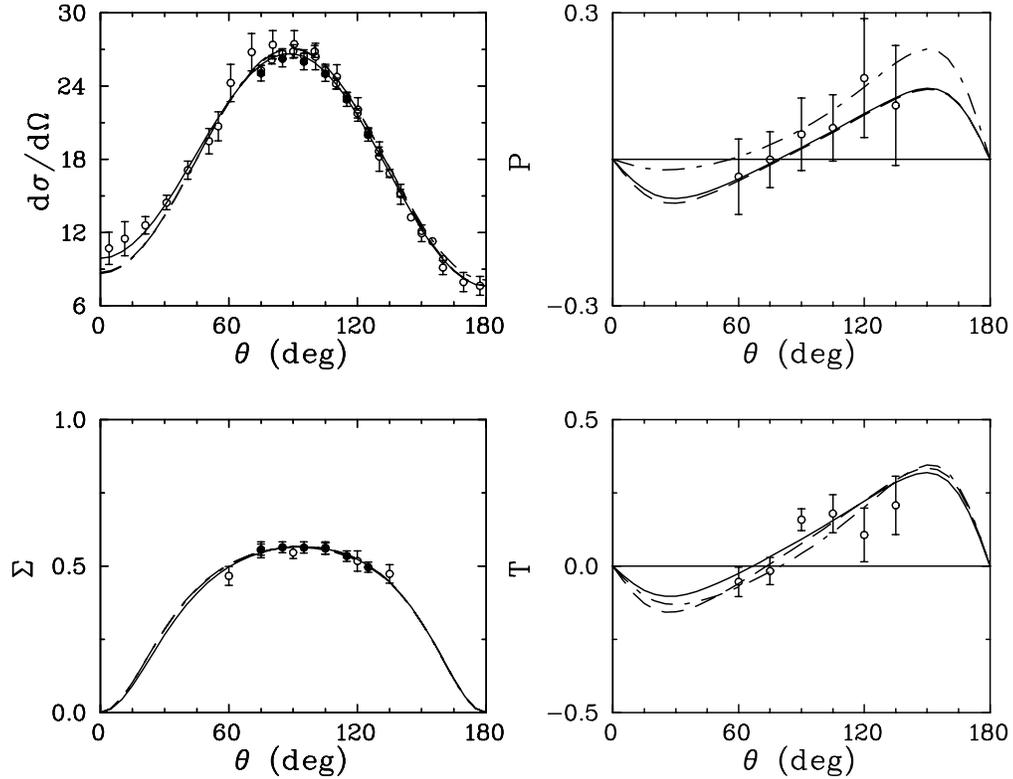}
\caption{\label{fig:g35} Fits to $\pi^0 p$ type-$S$ observables at
350 MeV; Fit1 (solid), Ref.~\cite{grushin} (dashed), Ref.~\cite{grushin}
(Born for $\ell$ $>$ 1) dot-dashed. Post-1990 data~\cite{beck} (solid),
pre-1990 data~\cite{data} (open) symbols.
}
\end{figure}

\begin{figure}
\includegraphics[ width=350pt, keepaspectratio, angle=90]{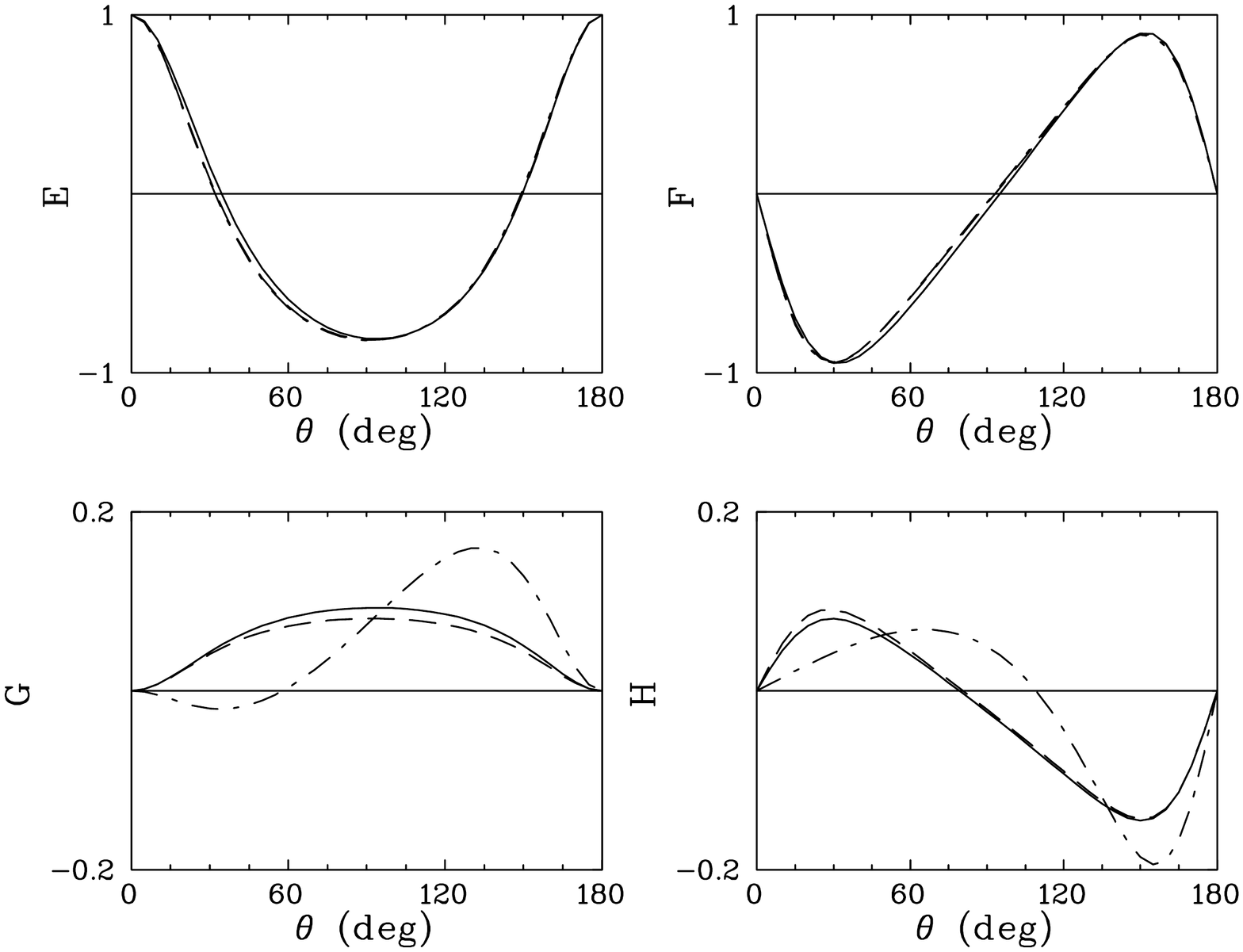}
\caption{\label{fig:gaf} Prediction of $\pi^0 p$ beam-target observables at
350 MeV; Fit1 (solid), Ref.~\cite{grushin} (dashed), Ref.~\cite{grushin}
(Born for $\ell$ $>$ 1) dot-dashed.
}
\end{figure}

Here, neglecting higher$-\ell$ multipoles leaves an undetermined overall phase. In a
fit from Ref.~\cite{grushin}, this phase was determined by setting Re $M_{1+}$ to a
fixed value. In Fit1, we have fixed
instead Im $M_{1+}$. In Fit 2, a value from the SAID energy-dependent fit
was assumed for $M_{1+}^{\pi^+ n}$ and
a parameter $\alpha$ was fitted using Watson's theorem~\cite{watson},
\begin{equation}
M_{1+}^{\pi^0 p} \; = \; \alpha e^{i \delta_{33}} +  \frac{1}{\sqrt{2}} M_{1+}^{\pi^+ n}
\end{equation}
with $\delta_{33}$ being the $P_{33}$ phase from elastic $\pi N$ scattering.
Two values for
$\alpha$ were found, positive and negative, the positive value being chosen above. In Fit1,
a second solution with Re $M_{1+}$ positive was also found. Fits 1 and 2 produce exactly
the same observables, leading to transversity amplitudes with fixed relative phases
but different overall phases.
The result labeled Fit 3 was obtained by conjugating the roots of the complex polynomials
for each transversity amplitude~\cite{omelaenko} from Fit 1. 
This is a symmetry of the type-$S$ observables
and half of the double-polarization quantities. The resulting solution is therefore not related
to Fits 1 and 2 by a rotation of the multipoles. As fits 1 through 3 give identical results for
type-$S$ observables, further information is required to select the correct solution. If
the multipoles of Fit 3 are rotated to have a phase for $M_{1+}^{\pi^0 p}$ matching Fit 2, the
resulting values for $E_{1+}^{\pi^0 p}$ will not combine, via Eq.~4, to give the proper 
phase for $E_{1+}^{3/2}$. 
In Ref.~\cite{grushin}, the neutral and charged pion results
were combined in an isospin analysis assuming that the $E_{1+}^{3/2}$ and $M_{1+}^{3/2}$
amplitudes had the same phase, without fixing this to be the phase from $\pi N$ elastic
scattering. However, given the sizeable errors found for $E_{1+}$, a direct application
of Watson's theorem seemed more effective.

In Fig.~4, the fit from Ref.~\cite{grushin} is compared to Fit1 and data for type-$S$
observables. We also show the effect of adding the MAID Born contribution,
for waves with $\ell > 1$, to the Grushin multipoles in Table III. The effect
is minimal except for $P$, which changes significantly, but not outside the large
uncertainties of these data. These comparisons are carried over to the
beam-target set in Fig.~5. As in Fig.~3, for $\pi^+ n$, the quantities $E$ and $F$
are quite stable, while $G$ and $H$ change significantly with the addition of
the higher$-\ell$ contributions, given by vector-meson exchange. The curves with
this addition look more like the MAID result. Fits 1 and 3 give identical results for 
$E$ and $H$, but have opposite signs for $G$ and $F$.

Somewhat different results for $P$ and $G$ were
also found when the $\Sigma$ data~\cite{beck}, used in the fit, were replaced by a measurement
with wider angular coverage~\cite{leukel}. In addition, preliminary measurements of a quantity 
proportional to $G$ appear to have a shape unlike that 
predicted by Fit1~\cite{sandorfi}. Precise measurements of $P$ and $G$ would clearly help
to stabilize the fit. 

\section{Conclusions}

We have reexamined the extraction of pion photoproduction 
multipoles from type-$S$ data with minimal model
input. In the process, we have suggested that deviations from recent fits covering
the resonance region may be due to problems in the database. This study should also give
some qualitative guidance to those who plan to extract multipoles from the 
present generation of polarized photoproduction experiments. The results given here
suggest that very precise data will be required for a reliable extraction of all but
the dominant multipoles. 
This is particularly evident it Table~II, where a sizeable change in Im $E_{0+}$ and
a wrong sign for Re $M_{1+}$ are linked to modest changes in the fit to $P$ data.

The procedure for $\pi^+ n$ photoproduction could be continued
up to higher energies, if the real high-$\ell$ multipole assumption remains
valid. For $\pi^0 p$, the existence of multiple solutions makes an isospin decomposition
more challenging. The use of Eq.~4 is also restricted to energies where the $P_{33}$ phase
is elastic. Finally, we note the 
possibility of accidental symmetries, generating solutions beyond those considered here.
This possibility was considered in Ref.~\cite{omelaenko}. 

\begin{acknowledgments}

This work was supported in part by the U.S. Department of
Energy Grant DE-FG02-99ER41110. The author thanks M. Paris and L. Tiator for 
discussions regarding the general amplitude reconstruction problem. 
\end{acknowledgments}
\eject

\eject

\end{document}